\documentclass[journal]{IEEEtran}
\usepackage{amsfonts,amsmath}
\usepackage{bm}
\usepackage{cuted}
\usepackage{cite}
\usepackage{xcolor}
\usepackage{algorithm}
\usepackage[noend]{algpseudocode}

\usepackage{graphicx}
\usepackage{subfigure}
\usepackage{float}
\usepackage{ifpdf}
\ifpdf
  \usepackage{epstopdf}
\fi
\ifCLASSINFOpdf
\else
\fi

\begin{document}
%
\title{Structural Balance of Complex Weighted Graphs\\ and Multi-partite Consensus}
%
%
%

\author{Honghui Wu,
Ahmet Taha Koru,
Guanxuan Wu,
Frank L. Lewis,
Hai Lin
\thanks{Honghui Wu and Hai Lin are with the Department
of Electrical Engineering, University of Notre Dame, Notre Dame,
IN, 46556 USA (e-mail: hwu9@nd.edu; hlin1@nd.edu)}
\thanks{Ahmet Taha Koru and Frank L. Lewis are with the UTA Research Institute, University of Texas at Arlington, Fort Worth, TX 76118 USA (e-mail: ahtakoru@gmail.com; lewis@uta.edu)}
\thanks{Guanxuan Wu is with the Department of Computer Science and Engineering, University of Texas at Arlington, Arlington, TX 76019 (e-mail: gxw6804@mavs.uta.edu)}
}

\maketitle

\begin{abstract}
The structural balance of a signed graph is known to be necessary and sufficient to obtain a bipartite consensus among agents with friend-foe relationships. In the real world, relationships are multifarious, and the coexistence of different opinions is ubiquitous. 
We are therefore motivated to study the multi-partite consensus problem of multi-agent systems, for which we extend the concept of structural balance to graphs with complex edge weights. It is shown that the generalized structural balance property is necessary and sufficient for achieving multi-partite consensus. 
Examples of multi-opinion social networks are illustrated as a real-world application of the proposed theoretical results.
\end{abstract}


%
\IEEEpeerreviewmaketitle

\section{Introduction}
%
%
%
%
Structural balance is a concept that originated from social psychology \cite{wasserman1994social}. It characterized the relationships in a friend-foe social network, which is usually modeled by a signed graph. If the agents over a signed graph are divided into two parties based on opposite opinions, where the edge weights between agents in the same party are positive (representing collaborative interaction), whereas those joining agents in different parties are negative (representing antagonistic interaction), then the signed graph and the corresponding social network are called \textit{structurally balanced}. The concept of bipartite consensus was proposed to capture the collective behaviors of agents over a signed graph, where the agents converge to a value with the same modulus but opposite signs. It was proved in \cite{altafini2012consensus} that structural balance is both necessary and sufficient to achieve bipartite consensus. Since then, bipartite consensus has become an active research area, see e.g., \cite{amirkhani2022consensus,zhang2014bipartite,bhowmick2019leader,xu2022observer}.


In reality, however, a social network is more than a dichotomy, and relationships are more complicated than mere friend-foe. For example, in international relationships, two countries have an adversarial bond due to opposite ideologies; they can still achieve certain agreements via a third country's mediation. The third country can retain neutrality without gravitating towards friendship or enmity with either nation. In this scenario, the traditional concept of structural balance falls short of describing the relationships, and bipartite consensus fails to encapsulate the collective behaviors. 

We are, therefore, motivated to study the multi-partite consensus problem. To depict the multifaceted relationships among agents, we introduce complex edge weights and generalize the social network models into \textit{complex weighted graphs}. The challenge lies in determining how to depict the relationships between agents using complex edge weights and establishing the conditions under which multi-partite consensus can be achieved. 

The contribution of this paper is threefold. First, we formally define the concept of structural balance of complex weighted graphs, study the properties of the graphs, {\color{black}and provide an algorithm to check structural balance}. Secondly, it is proved that structural balance is a necessary and sufficient condition for multi-agent systems to achieve multi-partite consensus. Thirdly, social networks with multifaceted relationships are modeled by complex weighted graphs with signatures representing agents' opinions. 


\noindent\textbf{Notations} This paper uses the following notations. Let $\mathbb{N}$, $\mathbb{R}$, and $\mathbb{C}$ be the set of natural numbers, real numbers, and complex numbers, respectively. Denote the imaginary unite $\imath=\sqrt{-1}$, then a complex number $z\in\mathbb{C}$ can be represented in the following forms: (1) $z=a+ \imath b$, where $a=Re(z)\in\mathbb{R}$ and $b=Im(z)\in\mathbb{R}$ are the real part and imaginary part of $z$, respectively; (2) $z=|z| e^{\imath \theta}$; (3) $z=|z|\angle{\theta}$, where $|z|$ is the modulus and $\theta\in(-\pi,\pi]$ is the argument of $z$. Note that $|z|=\sqrt{a^2+b^2}$, $\theta=\tan^{-1} (b/a)$. Denote $\bm{1}$ and $\bm{1}_n$ the vector with all elements as 1 with appropriate length and length of $n$, respectively. A matrix is called non-negative if all its entries are non-negative real numbers. In this paper, the terms node and agent are used interchangeably, referring to the same concept.

\section{Complex Weighted Graphs}
\label{section:complex_weighted_graphs}

Recall that a graph is denoted by $\mathcal{G}(A)=(\mathcal{V},\mathcal{E})$, where $\mathcal{V}=\{v_1,\dots,v_N\}$ is the node (or vertex) set, and $\mathcal{E}=\mathcal{V}\times\mathcal{V}$ is the edge set. 
{\color{black} To depict the multifaceted relationships
among agents, we allow the edge weights to be complex numbers and accordingly call them \textit{complex weighted graphs}.  To be precise, the elements in the adjacency matrix $A$ of a complex weighted graph $\mathcal{G}(A)$ could be a complex number, namely $a_{ij}\in\mathbb{C}$, $i \neq j$ in  $A=[a_{ij}]_{N \times N}$.} Note that $a_{ij}\neq 0$ when there is an edge from node $v_j$ to node $v_i$, i.e., $(v_j,v_i)\in\mathcal{E}$. For an edge $(v_j,v_i)$, we call node $v_j$ the parent node and $v_i$ the child node. Define the row connectivity matrix $D=diag(d_1,\dots,d_N)$, where $d_i=\sum_{j\in\mathcal{N}_i}|a_{ij}|$, for $i=1,\dots,N$, is the in-degree of node $i$, and $\mathcal{N}_i=\{ j| a_{ij} \neq 0 \}$ is the neighbor set of node $i$. Then the Laplacian of $\mathcal{G}(A)$ is defined as $L=D-A$.

The following notations are adopted here. We say that there is a \textit{directed path} from node $v_{i_1}$ to node $v_{i_n}$, denoted by $\mathcal{DP}_{(v_{i_1},v_{i_n})}$, if there is a sequence of nodes $v_{i_1},\dots,v_{i_n}$ such that $(v_{i_{l-1}},v_{i_{l}})\in\mathcal{E}$ for $l=2,\dots,n$. When we do not care about the direction of the edges, a \textit{weak path} can be defined from node $v_{i_1}$ to node $v_{i_n}$, denoted by $\mathcal{WP}_{(v_{i_1},v_{i_n})}$, if there is a sequence of nodes $v_{i_1},\dots,v_{i_n}$ such that $(v_{i_{l-1}},v_{i_{l}})\in\mathcal{E}$ or $(v_{i_{l}},v_{i_{l-1}})\in\mathcal{E}$ for $l=2,\dots,n$. Accordingly, a \textit{weak cycle} (\textit{directed cycle}) is a weak path (direct path) that begins and ends at the same node. We denote a directed cycle and a week cycle containing node $v_{i}$ by  $\mathcal{DC}_{(v_{i})}\equiv\mathcal{DP}_{(v_{i},v_{i})}$ and $\mathcal{WC}_{(v_{i})}\equiv\mathcal{WP}_{(v_{i},v_{i})}$, respectively. Note that a directed path (directed cycle) is a weak path (weak cycle) but not vice versa.

We next extend the structural balance concept from signed graphs (see \cite{altafini2012consensus,zhang2014bipartite}) to complex weighted graphs. To facilitate the definition of structural balance, we define signatures $\theta_1,\dots,\theta_N\in(-\pi,\pi]$ of the nodes $v_1,\dots,$ $v_N$ as the indicators stating which partite a node belongs to, i.e., the nodes with the same signatures are in the same partite(s).




{\color{black}
\textbf{\emph{Definition 1 (Structural Balance):}}
A graph $\mathcal{G}(A)$ is said \emph{structurally balanced} if the nodes $v_1,\dots,$ $v_N$ can be assigned with signatures $\theta_1,\dots,\theta_N\in(-\pi,\pi]$, respectively, such that all the entries of its adjacency matrix $A=[a_{ij}]_{N\times N}$ satisfies $a_{ij}\equiv |a_{ij}|\angle\theta_{ij} =|a_{ij}|\angle (\theta_i-\theta_j)$.
}

 We use a social network to illustrate the meaning of signature and the definition of structural balance. In a social network modeled by a structurally balanced $\mathcal{G}(A)$, the signatures $\theta_i$ and $\theta_j$ {\color{black}signify the values and beliefs} of agents $v_i$ and $v_j$ respectively. The argument $\angle\theta_{ij}=\angle (\theta_i-\theta_j)$ of the edge weight $a_{ij}$ represents the relationship between the two agents. In other words, the argument of the edge weight $\angle\theta_{ij}$ between two nodes equals the difference between their signatures, which implies the relationship between two agents depends on  {\color{black} the difference of their values and beliefs}.
We have the following result for a structurally balanced graph.





\textbf{\emph{Lemma 1:}} The followings are equivalences:

(1) Complex weighted graph $\mathcal{G}(A)$ is structurally balanced.

(2) Zero is a eigenvalue of $L$ with eigenvector {\color{black}$\bm{\zeta}=[1\angle\theta_1,\dots,1\angle\theta_N]^T$}.

(3) $D_\zeta:=diag(\bm{\zeta})$ such that $\widehat{A}=D_\zeta^{-1}AD_\zeta$ is nonnegative and $\widehat{A}=[|a_{ij}|]_{N \times N}$.

(4) $D_\zeta:=diag(\bm{\zeta})$ such that {\color{black}$\widehat{L}=D_\zeta^{-1}LD_\zeta$} has a zero eigenvalue with {\color{black}an} eigenvector being $\bm{1}$, {\color{black}where $\widehat{L}=D-\widehat{A}$.}

\emph{Proof:} 
{\color{black}(1)$\Leftrightarrow$(2): $\forall i,j \in [1,N], \text{ } a_{ij} = |a_{ij}|\angle(\theta_i - \theta_j) \Leftrightarrow \forall i\in[1,N], \text{ } \sum_{j\in\mathcal{N}_i}|a_{ij}|=\sum_{j\in\mathcal{N}_i}a_{ij}\cdot\big(1\angle(\theta_j-\theta_i)\big) \Leftrightarrow \forall i\in[1,N], \text{ } d_i\angle\theta_i-\sum_{j\in\mathcal{N}_i}a_{ij}\cdot(1\angle\theta_j)=0 \Leftrightarrow L\bm{\zeta}=0$.}

(1)$\Leftrightarrow$(3): (Sufficiency) According to the {\color{black}definition 1}, it shows $\widehat{A}=D_\zeta^{-1}AD_\zeta$ where $\widehat{A}=[|a_{ij}|]_{N \times N}$ by expanding the matrices. (Necessity) Given  $\widehat{A}=[|a_{ij}|]_{N \times N}$, let $A=D_\zeta \widehat{A} D_\zeta^{-1}$, then $A$ is the adjacency matrix of structurally balanced graph $\mathcal{G}(A)$ with signature $\theta_1,\dots,\theta_N$ from the eigenvector $\bm{\zeta}$.

(2)$\Leftrightarrow$(4): $L\zeta=0 \Leftrightarrow LD_\zeta\bm{1}=0 \Leftrightarrow D_\zeta^{-1}LD_\zeta\bm{1}=0 \Leftrightarrow \widehat{L}\bm{1}=0$. \hfill $\Box$

{\color{black}Consider a structurally balanced $\mathcal{G}(A)$.} Lemma 1 (2) gives us a closed-form solution of the eigenvector $\bm{\zeta}$: for an entry of $\bm{\zeta}$, the argument is the signature of the corresponding node, and the modulus is 1.
Lemma 1 (3) and (4) show that the Laplacian $L$ and adjacency matrix $A$ of a structurally balanced $\mathcal{G}(A)$ can be respectively transformed into the Laplacian $\widehat{L}$ and adjacency matrix $\widehat{A}$ of graph $\mathcal{G}(\widehat{A})$ with nonnegative real edge weights. This implies that abundant tools from graph theory established on the nonnegative real graphs can be applied to structurally balanced $\mathcal{G}(A)$ once we obtain the similarity transformation matrix $D_\zeta$ according to $\bm{\zeta}$. If the $\mathcal{G}(A)$ is constructed by assigning signatures, then $\bm{\zeta}$ is directly obtained from Lemma 1 (2).

{\color{black}\textbf{\emph{Remark 1:}}  If we are to construct a graph, we can first assign the nodes with signatures and then define the edge weights to ensure structural balance.} However, in most cases, a graph is given to us rather than constructed by us, and only the connection topology and edge weights are defined. {\color{black}In other words, only the adjacency matrix is known, and the signatures are not given.} We, therefore, ask how to check whether a graph $\mathcal{G}(A)$ is structurally balanced and how to solve the eigenvector $\bm{\zeta}$ without prior knowledge of signatures.










To answer these questions, we propose a spanning tree-based method to assign signatures.
Recall that a spanning tree is a graph with a directed path from at least one single node, called the root, to all other nodes without forming any weak cycles; and a graph $\mathcal{G}(A)=(\mathcal{V},\mathcal{E})$ is said connected if it contains a spanning tree as a subgraph $\mathcal{G}'(A)=(\mathcal{V},\mathcal{E}')$, where $\mathcal{E}'\subseteq \mathcal{E}$. 






\textbf{\emph{Lemma 2:}} A spanning tree is structurally balanced.

\emph{Proof:} Suppose that the edge weight from a parent node $v_{i_1}$ to a child node $v_{i_2}$ equals $|a_{i_2i_1}|\angle\theta_{i_2i_1}$. Beginning with the root, assign the parent node with signature $\theta_{i_1}$, then the signature of the child node is assigned by $\theta_{i_2}:=\theta_{i_1}+\theta_{i_2i_1}$. Do the same by treating the children of the root as parents, and so on, until all nodes are assigned signatures. This is always feasible because every node, except the root, has and only has one parent node on a spanning tree. Then for any edge weight on the spanning tree, it always satisfies  $a_{i_2i_1}\equiv|a_{i_2i_1}|\angle\theta_{i_2i_1}=|a_{i_2i_1}|\angle(\theta_{i_2}-\theta_{i_1})$. \hfill $\Box$

The above proof tells us that to solve $\bm{\zeta}$ from a connected and structurally balanced graph, one can begin with finding a spanning tree and then perform signature assignment to the nodes according to the proof for Lemma 2. Note that $b\cdot\bm{\zeta}$ for any constant $b\in\mathbb{C}$ is also an eigenvector corresponding to eigenvalue 0 of the Laplacian. In other words, the solution of $\bm{\zeta}$ is not unique. 
This is because one can find a different spanning tree or assign the nodes with different signatures satisfying the constraints by the arguments of edge weights. 
Fortunately, if the graph is connected, the eigenvector is definite once an element in the eigenvector is defined because the multiplicity of eigenvalue zero is one for connected graphs. The following corollary captures this property.


\textbf{\emph{Corollary 1:}} Structurally balanced $\mathcal{G}(A)$ is connected if and only if 
zero is a \textit{simple eigenvalue} of $L$ with eigenvector $\bm{\zeta}$.


\emph{Proof:} Note that a nonnegative graph $\mathcal{G}(\widehat{A})$ is connected if and only if the zero eigenvalue of its Laplacian $\widehat{L}$ is of multiplicity of one (refer Theorem 2.1 in \cite{lewis2013cooperative}). Then the corollary is proved with Lemma 1. \hfill $\Box$

Inspired by Lemma 2 and Corollary 1, we obtain the following algorithm to check the structural balance of a graph and solve the eigenvector $\bm{\zeta}$ if the graph is structurally balanced.


\begin{algorithm}
\caption{Check structural balance and Solve $\bm{\zeta}$}
\begin{algorithmic}[1]

{\color{black}
\State \textbf{Input:} $\theta_{ij}, \forall i,j=1,\dots,N\text{ and }i\neq j$ 

a spanning tree $\mathcal{G}'(A)=(\mathcal{V},\mathcal{E}')$

\color{black}
\State $\theta_r\gets0$ \Comment{$\theta_r$ is the signature of the root node}

\For{$(v_j,v_i)\in\mathcal{E}'$} 
\State $\theta_{i}\gets\theta_j+\theta_{ij}$
\EndFor

\color{black}
\For{$(v_j,v_i)\notin\mathcal{E}'$} 
\If{$\theta_{ij}!=\theta_{i}-\theta_j$} 
 
 \Return The graph is structurally unbalanced; break; 

\EndIf
\EndFor
}






\State \Return The graph is structurally balanced,

and $\bm{\zeta}=[1\angle\theta_1,\dots,1\angle\theta_N]$

\end{algorithmic}
\end{algorithm}





In Algorithm 1, the inputs are all the arguments of edge weights and a spanning tree. It begins with setting the signature of the root node $\theta_r:=0$. Then, for each edge $(v_j,v_i)$ on the spanning tree, define the signature of the child node to be the sum of the signature of the parent node and the argument of the edge, i.e., $\theta_i:=\theta_j+\theta_{ij}$, beginning from the root node to all other nodes. For each edge $(v_j,v_i)$ not on the spanning tree, check whether the argument $\theta_{ij}$ satisfies $\theta_{ij}=\theta_i-\theta_j$. If they all satisfy, then the graph is structurally balanced, and the eigenvector $\bm{\zeta}$ is obtained with the signatures. 

Note that Algorithm 1 is for a connected graph. For a non-connected graph, the algorithm can still be applied by first decomposing the graphs into two or more connected sub-graphs.
Alternatively,  we provide another way to check whether a graph is structurally balanced, which can be directly applied to a connected or non-connected graph. For this, we first extend the concept of consistency of weak cycle in \cite{lou2014distributed,yaghmaie2017multiparty} from unit modulus graphs to general graphs. 

\textbf{\emph{Definition 2 (Consistent Cycle):}} A weak cycle containing node $v_{i_0}$ is said to be \textit{consistent} if
\begin{equation}\label{consistentcycle}
    \prod_{i\in\{i|(v_i,v_{i+1})\in\mathcal{WC}_{(v_{i_0})}\}}\omega_i\in\mathbb{R},
\end{equation}
where $\omega_i=a_{i,i+1}$ if $(v_{i},v_{i+1})\in\mathcal{E}$, and $\omega_i=a_{i,i+1}^{-1}$ if $(v_{i+1},v_{i})\in\mathcal{E}$, $a_{i+1,i}$ and $a_{i,i+1}$ are the edge weight of $(v_{i},v_{i+1})$ and $(v_{i+1},v_{i})$, respectively.


Intuitively, a consistent cycle requires that the sum of the arguments of the edges in the clockwise direction minus the sum of those in the counterclockwise direction equals zero.

\textbf{\emph{Lemma 3:}} For a graph containing one or more weak cycle(s), the graph is structurally balanced if and only if all the weak cycles are consistent.

 \emph{Proof:} Assume the graph contains a spanning tree denoted as $\mathcal{G}'(A)=(\mathcal{V},\mathcal{E}')$ where $\mathcal{E}'\subseteq\mathcal{E}$. Assign the nodes with signatures according to the proof for Lemma 2. Without loss of generality, denote the signatures for nodes $v_i$ and $v_j$ as $\theta_i$ and $\theta_j$, respectively. It is already proven in Lemma 2 that for any edges on the spanning tree, the edge weights satisfy $a_{i_2i_1}\equiv|a_{i_2i_1}|\angle\theta_{i_2i_1}=|a_{i_2i_1}|\angle(\theta_{i_2}-\theta_{i_1})$. Now we prove Lemma 3 as follows by showing that for any edge $(v_i,v_j)\notin\mathcal{E}'$, (i.e., any edges on the weak cycle but not on the spanning tree), the edge weight satisfies $a_{ij}=|a_{ij}|\angle (\theta_i-\theta_j)$ if and only if \eqref{consistentcycle} holds:


\begin{small}
\begin{equation*}\label{consistentproof}
\begin{split}
    &\prod\omega_i=a_{i_k i_{k+1}}\dots a_{i_{k+n} i}\cdot a_{ij}^{-1} \cdot (a_{j_k j_{k+1}}\dots a_{j_{k+m} j})^{-1} \\
    &=|a_{i_k i_{k+1}}\dots a_{i_{k_n} i}|\angle[(\theta_{i_{k+1}}-\theta_{i_k})+\cdots+(\theta_{i}-\theta_{i_{k+n}})]\cdot a_{ij}^{-1}\cdot\\
    & \big(|a_{j_k j_{k+1}}\dots a_{j_{k+m} j}|)\angle[(\theta_{j_{k+1}}-\theta_{j_k})+\cdots+(\theta_{j}-\theta_{j_{k+n}})]\big)^{-1}\\
    &=a_{ij}^{-1} \cdot \frac{|a_{i_k i_{k+1}}\dots a_{i_{k_n} i}|}{|a_{j_k j_{k+1}}\dots a_{j_{k+m} j}|}\angle(\theta_i-\theta_j)
\end{split}
\end{equation*}
\end{small}
where $\theta_{i_k}=\theta_{j_k}$ because it is the signature of the same node $v_k$. From above equation, $\prod\omega_i\in\mathbb{R}$ if and only if $a_{ij}=|a_{ij}|\angle(\theta_i-\theta_j)$. \hfill $\Box$

Lemma 3 provides an intuitive way to check the structural balance of a graph. Compared with the results from \cite{dong2014complex,dong2016laplacian}, which check structural balance based on the Hermitian part of adjacency matrix $A$, our results also provide a solution of eigenvector $\bm{\zeta}$.

{\color{black}\textbf{\emph{Remark 3:}} The computation complexity of Algorithm 1 is $O(N^2)$ as it traverses the edges. Checking structural balance with Lemma 3 is NP-hard because a graph can have exponentially many weak cycle as the number of nodes and edges increases. Similarly, checking structural balance with the results from \cite{dong2014complex,dong2016laplacian} is also NP-hard because they also check the (weak) cycles in a graph. Therefore, this paper provides Algorithm 1 as a practical method to check structural balance of large and dense graphs.

Note that Algorithm 1 uses a spanning tree as input, and there are many mature algorithms to find a spanning tree within $O(N^2)$. One classical method is based on strongly connected components (SCCs) by Tarjan \cite{tarjan1972alg}. Given a graph for spanning tree look-up, it first finds out all the strongly connected subgraphs, a.k.a., strongly connected components (SCCs). A graph is said strongly connected if every node is reachable from every other node.
Then it condenses the graph by regarding each SCC as a vertex and the graph becomes a directed acyclic graph (DAG).  With the DAG, one can easily find a spanning tree by choosing a vertex has no incoming edges in the DAG. The chosen vertex is either a single node or a SCC. If it is single node, then it is a root node from which one can easily generate a spanning tree along the DAG. If it is a SCC, then any node in the SCC can be a root node from which generates a spanning tree along the DAG. The above processes is bounded by $O(N^2)$.



}





\section{Multi-partite Consensus}

 It turns out that the structural balance is necessary and sufficient for multi-partite consensus. Before we prove this claim, we need to specialize the concept of structural balance into multi-partite structural balance.




\textbf{\emph{Definition 3 (Multi-partite Structural Balance):}} Graph $\mathcal{G}(A)=(\mathcal{V},\mathcal{E})$ is said to be  \emph{multi-partite structurally balanced} if it admits a $k$-partition of the nodes $\mathcal{V}_1,\dots,\mathcal{V}_k$, with $\mathcal{V}_1\cup\dots\cup\mathcal{V}_k=\mathcal{V}$, $\mathcal{V}_p\cap\mathcal{V}_q=\O$, where $k\in\mathbb{N}_+$, $p,q\in\{1,\dots,k\}$, and $p\neq q$, such that $a_{ij}=|a_{ij}|$ if $v_i,v_j\in\mathcal{V}_p$, and $a_{ij}=|a_{ij}|\angle (\theta_p-\theta_q)$ if $v_i\in\mathcal{V}_p, v_j\in\mathcal{V}_q$, where $\theta_p\neq\theta_q$, $\theta_p,\theta_q\in(-\pi,\pi]$ are the signature of nodes in partition $\mathcal{V}_p$ and $\mathcal{V}_p$, respectively.

{\color{black}\textbf{\emph{Remark 4:}} After assigning the nodes with signatures using Algorithm 1, one is ready to verify multi-partite structural balance by the above definition. Since it is a special case of Definition 1, Lemma 1 holds for multi-partite structurally balanced $\mathcal{G}(A)$, where accordingly the eigenvector of zero eigenvalue becomes $\bm{\zeta}=[1\angle\theta_i]_{N\times 1}$ for $i=1,\dots,n$, where $\theta_i=\theta_p$ if node $v_i\in\mathcal{V}_p$.}

{\color{black} 
The following introduces multi-partite consensus to depict the opinion dynamics of a structurally balanced social network, which generalizes the bi-polarization of opinions over signed graphs (bipartite graphs) to multi-polarization over complex weighted graphs (multi-partite graphs).


}

\textbf{\emph{Definition 4 (Multi-partite consensus):}} A system of $N$ agents with the $i$-th agent $v_i$ modeled by 
\begin{equation}
    \dot{x}_i(t)=f_i(x_i,t),\quad i=1,\dots,N,
\end{equation}
where $x_i\in\mathbb{C}^n$ is the state vector and $f_i(\cdot,\cdot):\mathbb{C}^n\times\mathbb{R}\to\mathbb{C}^n$ is the agent's dynamics, achieves \textit{multi-partite consensus} if 
\begin{equation}\label{GlobalmultipartiteConsensus}
    \lim_{t\to\infty} x(t) = c \cdot \bm{\zeta} \otimes \bm{1}_n,
\end{equation}
where $x(t)=[x_1(t),\dots,x_N(t)]^T\in\mathbb{C}^{nN}$, $\bm{\zeta}=[1\angle\theta_i]_{N\times 1}$ with $\theta_i=\theta_p$ when agent $v_i\in\mathcal{V}_p$, and $c\in\mathbb{C}$ is a complex constant. 

A multi-agent system achieving multi-partite consensus means that the agents converge to complex values with the same modulus but different arguments defined by the partitions. 
We first consider a first-order multi-integrator system to formulate the main theorem and then generalize it into general-order linear multi-agent systems.

\textbf{\emph{Theorem 1:}} A first-order continuous-time multi-integrator system of $N$ agents
\begin{equation}\label{CTintegrator}
    \dot{x}_i(t)=u_i(t),\quad i=1,\dots,N,
\end{equation}
where $x_i(t)\in\mathbb{C}$ and $u_i(t)\in\mathbb{C}$ are the state and input of $i$-th agent, respectively, with the following protocol
\begin{equation} \label{CTu1}
\begin{split}
    u_i(t)&=-\sum_{j\in\mathcal{N}_i}\Big(|a_{ij}| x_i(t) -a_{ij} x_j(t) \Big)\\
    &=-\sum_{j\in\mathcal{N}_i}|a_{ij}|\Big( x_i - \big(1\angle\theta_{ij}\big) \cdot x_j \Big),
\end{split}
\end{equation}
where $a_{ij}\equiv|a_{ij}|\angle\theta_{ij}\in\mathbb{C}$ is the edge weight of graph $\mathcal{G}(A)$, achieves multi-partite consensus (i.e., $\lim_{t\to\infty} x(t) = c \cdot \bm{\zeta}$), if and only if, graph $\mathcal{G}(A)$ is connected and multi-partite structurally balanced.


{\color{black}
\emph{Proof:} Rewrite the system \eqref{CTintegrator} along the protocol \eqref{CTu1} into a global form $\dot{x}(t)=-Lx(t)$,
where $x(t)=[x_1(t),\dots,x_n(t)]^T$. Note that $\widehat{L}=D_\zeta^{-1}LD_\zeta$ if and only if $\mathcal{G}(A)$ is structurally balance. By change of coordinates $z(t)=D_\zeta^{-1}x(t)$, one has $\dot{z}(t)=-\widehat{L}z(t)$.
Since $\widehat{L}$ is the Laplacian of nonnegative $\mathcal{G}(\widehat{A})$ with the same connectivity of $\mathcal{G}(A)$, $\widehat{L}$ has a simple eigenvalue zero with eigenvector $\bm{1}$ if and only if $\mathcal{G}(A)$ is connected. Moreover, by Gershgorin circle theorem, all non-zero eigenvalues of $\widehat{L}$ have positive real parts. These imply $\lim_{t\to\infty}z(t)=\bm{1}w_1^Tz(0)$, where $w_1^T$ is the left eigenvector of zero eigenvalue of $\widehat{L}$. Noting that $D_\zeta\bm{1}=\bm{\zeta}$ and $w_1^Tz(0)=w_1^TD^{-1}_\zeta x(0)$ is a scalar, i.e., a complex constant, one has $\lim_{t\to\infty}x(t)=(w_1^TD^{-1}_\zeta x(0))\cdot\bm{\zeta}$.  \hfill $\Box$
}

\color{black} 
\textbf{\emph{Remark 6:}} The state $x_i(t)\equiv|x_i(t)|\angle\alpha_i(t)\in\mathbb{C}$ can be interpreted as an opinion expressed by agent $v_i$ at time $t$, where the magnitude $|x_i(t)|$ represents the strength or intensity and the argument $\angle\alpha_i(t)$ represents the stance of the opinion. Then, multi-partite consensus means, with the communication protocol \eqref{CTu1}, all the opinions by the agents reach the same intensity of strength with different stances, and the stances of opinions are multi-polarized corresponding to the agents' values and beliefs stored in $\bm{\zeta}=[1\angle\theta_i]_{N\times 1}$ for $i=1,\dots,N$. 

\textbf{\emph{Remark 7:}} Examples to further illustrate the interpretation in social networks will be shown in section III in this paper. Apart from social networks, multi-partite consensus and complex-valued Laplacian are also widely applied to engineering scenarios, such as target surrounding \cite{lou2014distributed}, formation control \cite{yaghmaie2017multiparty,ranjbar2020event,wang2022attack}, and power system \cite{halihal2022estimation}.

%





{\color{black}  It is obvious that ordinary consensus and bipartite consensus are special cases of multi-partite consensus by limiting the states in real numbers and the signatures of nodes in single value 0 and discrete values 0 and $\pi$, respectively. 
Also, the extension of multi-partite consensus control to multi-agent systems with more general linear dynamics is straightforward.

\textbf{\emph{Corollary 3:}} 
The following two problems are equivalent {\color{black}in terms of designing the control gain $K$. In other words, the same control gain $K$ solves the two problems}:
\begin{itemize}
    \item \textbf{\textit{Multi-partite Consensus Control:}} Design a control gain $K\in\mathbb{R}^{m\times n}$ such that a high-order multi-agent system of $N$ agents achieves multi-partite consensus over a connected structurally balanced graph $\mathcal{G}(A)$, with the $i$-th agent modeled by
\begin{equation}\label{LTIMASs}
    \dot{x}_i=\mathcal{A} x_i + \mathcal{B} u_i,\quad i=1,\dots,N,
\end{equation}
where $\mathcal{A}\in\mathbb{R}^{n\times n}$ and $\mathcal{B}^{n\times m}$ are system matrices and $(\mathcal{A},\mathcal{B})$ is controllable, $x_i\in\mathbb{C}^n$ and $u_i\in\mathbb{C}^m$ are the state vector and control input vector, respectively, with the following control input
\begin{equation}
\label{CTRLLAWs}
    u_i=-K\sum_{j\in\mathcal{N}_i}\Big(|a_{ij}| x_i -a_{ij} x_j \Big).
\end{equation}

\item \textbf{\textit{Ordinary Consensus Control:}} Design a control gain $K\in\mathbb{R}^{m\times n}$ such that a high-order multi-agent system of $N$ agents achieves ordinary consensus over a nonnegative graph $\mathcal{G}(\widehat{A})$, which relates to $\mathcal{G}(A)$ by {\color{black}$\widehat{A}=D_\zeta^{-1} A D_\zeta = [|a_{ij}|]_{n\times n}$}, with the $i$-th agent modeled by
\begin{equation}\label{LTIMASsZ}
    \dot{z}_i=\mathcal{A} z_i + \mathcal{B} \bar{u}_i,\quad i=1,\dots,N,
\end{equation}
where $\mathcal{A}$ and $\mathcal{B}$ are the same as \eqref{LTIMASs}, $z_i\in\mathbb{R}^n$ and $\bar{u}_i\in\mathbb{R}^m$ are the state vector and control input vector, respectively, with the following control input
\begin{equation}
\label{CTRLLAWz}
    \bar{u}_i=-K\sum_{j\in\mathcal{N}_i}|a_{ij}|\big( z_i -  z_j \big).
\end{equation}
\end{itemize}

{\color{black} Proof:} The overall closed-loop system due to \eqref{LTIMASs} and \eqref{CTRLLAWs} is $ \dot{x} = \left( I_N \otimes A - L \otimes BK \right) x $ where $ L $ is the Laplacian matrix defined in the beginning of Section \ref{section:complex_weighted_graphs} and $ x $ is defined in Definition 4. Similarly, the overall closed-loop system due to \eqref{LTIMASsZ} and \eqref{CTRLLAWz} is $ \dot{z} = \left( I_N \otimes A - \widehat{L} \otimes BK \right) z $ where $ \widehat{L} $ is defined in item (4) of Lemma 1 and $ z = \left[ z_1, \dots, z_N \right]^{T} $. From the same item of Lemma 1, we deduce that the closed-loop system matrices are similar due to the change of basis matrix $ (D_{\zeta} \otimes I_n) $  when $\mathcal{G}(A)$ is structurally balanced. Hence, they have the same closed-loop system eigenvalues. \hfill $\Box$

{\color{black}The above results for continuous-time systems can be easily extended to discrete-time systems. A discrete-time version of Theorem 1 is formulated as below.}

{\color{black}
\textbf{\emph{Corollary 4:}} A first order discrete-time multi-integrator system of $N$ agents
\begin{equation}\label{DTintegrator}
    x_i(k+1)=x_i(k)+u_i(k),\quad i=1,\dots,N,
\end{equation}
where $x_i(k)\in\mathbb{C}$ and $u_i(k)\in\mathbb{C}$ are the state and input of $i$-th agent, respectively, with the following protocol
\begin{equation} \label{DTu1}
\begin{split}
    u_i(k)&=-\kappa_i\sum_{j\in\mathcal{N}_i}\Big(|a_{ij}| x_i(k) -a_{ij} x_j(k) \Big)\\
    &=-\kappa_i\sum_{j\in\mathcal{N}_i}|a_{ij}|\Big( x_i - \big(1\angle\theta_{ij}\big) \cdot x_j \Big),
\end{split}
\end{equation}
where $a_{ij}\in\mathbb{C}$ is the edge weight of graph $\mathcal{G}(A)$, $ 0 < \kappa_i \leq 1/d_i$, $d_i=\sum_{j\in\mathcal{N}_i}|a_{ij}|$ is the in-degree of node $v_i$, achieves multi-partite consensus (i.e., $\lim_{k\to\infty} x(k) = c \cdot \bm{\zeta}$), if and only if, graph $\mathcal{G}(A)$ is connected and multi-partite structurally balanced.
}

{\color{black}Proof:  Let $ \kappa = diag(\kappa_1, \dots, \kappa_N) $, then the overall closed-loop system is $ x(k + 1) = \left(  I_N -  \kappa L  \right) x(k) $. Note that $\widehat{L}=D_\zeta^{-1}LD_\zeta$ if and only if $\mathcal{G}(A)$ is structurally balance. By change of coordinates $z(k)=D_\zeta^{-1}x(k)$, one has $ z(k + 1) =  \left(  I_N -  \kappa \widehat{L}  \right) z(k) $. From Lemma 3 in \cite{olfati2007consensus}, the matrix $ \left(  I_N -  \kappa \widehat{L}  \right) $ has a simple eigenvalue of 1, all other eigenvalues are in the unit circle. These imply $\lim_{k\to\infty}z(k)=\bm{1}w_1^Tz(0)$, where $w_1^T$ is the left eigenvector of the matrix $ \left(  I_N -  \kappa \widehat{L}  \right) $ corresponding to the eigenvalue of 1. Noting that $D_\zeta\bm{1}=\bm{\zeta}$ and $w_1^Tz(0)=w_1^TD^{-1}_\zeta x(0)$ is a scalar, i.e., a complex constant, one has $\lim_{k\to\infty}x(k)=(w_1^TD^{-1}_\zeta x(0))\cdot\bm{\zeta}$. } \hfill $\Box$

{\color{black} Compare to \eqref{CTu1} in Theorem 1 for continuous-time systems, the protocol \eqref{DTu1} for discrete-time systems includes an auxiliary gain $\kappa_i$ for stability. It is worth noting that $\kappa_i$ can be absorbed by a control gain $K$, and as a result, Corollary 3 also holds for discrete-time systems.}

\section{Application on Social Networks}



{\color{black}In reality, social networks are multifaceted, extending beyond simple friend-foe relationships. Complex weighted graphs (multi-partite graphs) provide a generalized model for social networks, encompassing the traditional signed graphs (bipartite graphs).}



\subsection{Modeling Social Networks with Complex Weighted Graphs}
{\color{black}
In a multifaceted social network modeled by a complex weighted graph, the signatures $\theta_1,\dots,\theta_N$ signify the values and beliefs of agents $v_1,\dots,v_N$, respectively. A directed edge represents a unilateral relationship between two agents. One example of the unilateral relationship is a person follows another person on Twitter. For another example, modeling a academic society, where the agents are scholars, an edge from node $v_j$ to node $v_i$, i.e., $(v_j,v_i)\in\mathcal{E}$, represents scholar $v_i$ is influenced by scholar $v_j$ such as by reading his/her publication. The edge weight of $(v_j,v_i)\in\mathcal{E}$ is defined as $a_{ij}\equiv |a_{ij}|\angle\theta_{ij}$. The magnitude $|a_{ij}|$ captures the strength of the influence from $v_j$ to $v_i$. The argument $\angle\theta_{ij}$ (also denoted by $\theta_{ij}$ for brevity) characterize the unilateral relationship between the two agents. If an edge from $v_i$ to $v_j$ also exists, i.e., $(v_i,v_j)\in\mathcal{E}$, then the relationship becomes mutual.

In a structurally balanced graph, the edge weight is $a_{ij}\equiv |a_{ij}|\angle\theta_{ij} =|a_{ij}|\angle (\theta_i-\theta_j)$ for any distinct pair of agents $v_i$ and $v_j$, which implies the relationship between any two agents is determined by the difference of their values and beliefs. Specifically, $\theta_{ij}=0$, $\theta_{ij}=\pm\frac{\pi}{2}$, and $\theta_{ij}=\pi$ indicate cooperative, neutral, and antagonistic relationships, respectively, arising from closely aligned, orthogonal, and directly opposed values and beliefs. $\theta_{ij}\in(0,\frac{\pi}{2})\cup(-\frac{\pi}{2},0)$ represents the relationship is partially cooperative resulting from their similar but not identical values and beliefs. Conversely, $\theta_{ij}\in(\frac{\pi}{2},\pi)\cup(-\pi,-\frac{\pi}{2})$ means the relationship is partially antagonistic resulting from their divergent but not completely opposite values and beliefs. If the relationship between agent $v_i$ and $v_j$ is mutual, then the argument of the edge weights must satisfy $\angle\theta_{ij}=\angle-\theta_{ji}$ to preserve structural balance. However, it is allowed that $|a_{ij}|\neq|a_{ji}|$. Note that the arguments with opposite signs indicate the same relationships. Structural balance requires a mutual relationship of two agents has the same level of "friendliness" or antagonism toward each other but allows for asymmetric strength of influence.}

\subsection{{\color{black}Opinion Dynamics of Structurally Balanced Networks}}
\begin{figure}[t!]
\centering
\subfigure[]{
\begin{minipage}[t]{0.15\textwidth}
\includegraphics[width=1\textwidth]{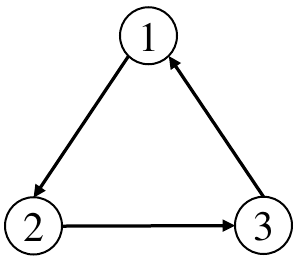}
\end{minipage}
}
\subfigure[]{
\begin{minipage}[t]{0.15\textwidth}
\includegraphics[width=1\textwidth]{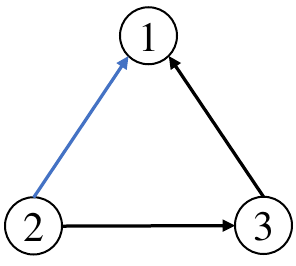}
\end{minipage}
}
\caption{Graph topology for the illustrative examples in social network. (a) Directed cycle with three nodes. (b) Weak cycle with three nodes.}
\end{figure}

{\color{black}Consider the discrete-time system \eqref{DTintegrator} and \eqref{DTu1} in Corollary 4. The state $x_i(k)$ and $x_j(k)$ represent opinions expressed by agent $v_i$ and agent $v_j$ at time $k$, respectively. $\big(1\angle\theta_{ij}\big) \cdot x_j$ represents how agent $v_i$ processes and perceives the opinion expressed by $v_j$, which is conditioned by their relationship stemming from the difference in their values and beliefs. 
Rewriting \eqref{DTintegrator} along with \eqref{DTu1}, it becomes
\begin{equation}\label{DTI1}
\begin{split}
x_i(k+1)
&=\big(1-\kappa_i d_i\big)x_i(k) + \kappa_i \sum_{j\in\mathcal{N}_i}|a_{ij}| \big(1\angle\theta_{ij}\big) \cdot x_j(k).
\end{split}
\end{equation}
It implies agent $v_i$'s {\color{black} opinion $x_i(k+1)$ is the sum of its weighted previous opinion $\big(1-\kappa_i d_i\big)x_i(k)$ and the processed and perceived opinions $\big(1\angle\theta_{ij}\big) \cdot x_j$ from its neighbors weighted by the strength of influence $|a_{ij}|$}. $\kappa_i$ is an auxiliary weight to stabilize the opinion dynamic. Specially, $\theta_{ij}=0$ implies agent $v_i$ agrees with and perceives the opinion from $v_j$ as it is. $\theta_{ij}=\pi$ implies agent $v_i$ completely disagrees with perceives the direct opposite of the opinion, as if the opposite of agent $v_j$ says is true. $\theta_{ij}=\pm\pi/2$ implies $v_i$ takes the opinion neutrally, neither aligning with nor opposing the opinion by $v_j$. Corollary 4 reveals that the above interactions among agents in a structurally balanced social network foster multi-partite consensus, where the opinions multi-polarize in terms of stances and reach the same level of intensity.
}





{\color{black}Structural balance also implies that the interaction among agent do not cause conflicts in perception. The following two examples show how it works, and more complicated cases can be analyzed similarly.}

\textit{Example 1:}
Consider the directed cycle in Fig 1 (a). For conciseness, let $|a_{ij}|=1$ and $\kappa_i=1/2$. According to \eqref{DTI1}, one has
\begin{align}
&x_1(k+1)=\frac{1}{2}\cdot x_1(k)+\frac{1}{2} \cdot \big(1\angle\theta_{13}\big)\cdot x_3(k)\label{x1k}\\
&x_3(k)=\frac{1}{2}\cdot x_3(k-1)+\frac{1}{2}\cdot\big(1\angle\theta_{32}\big) \cdot x_2(k-1)\label{x3k}\\
&x_2(k-1)=\frac{1}{2}\cdot x_2(k-2)+\frac{1}{2}\cdot\big(1\angle\theta_{21}\big) \cdot x_1(k-2)\label{x2k}
\end{align}
Substituting \eqref{x2k} into \eqref{x3k} and then \eqref{x3k} into \eqref{x1k}, one has
\begin{equation}\label{x1ksum}
\begin{split}
    x_1(k+1)=&\frac{1}{2}x_1(k) + \big(\frac{1}{2}\big)^2\big( 1\angle\theta_{13} \big)x_3(k-1)\\
    &+ \big(\frac{1}{2}\big)^3 \big( 1\angle(\theta_{13}+\theta_{32}) \big)x_2(k-2)\\
    &+ \big(\frac{1}{2}\big)^3 \big( 1\angle(\theta_{13}+\theta_{32}+\theta_{21}) \big)x_1(k-2)
\end{split}
\end{equation}

In \eqref{x1ksum}, $1\angle(\theta_{13}+\theta_{32}+\theta_{21})=1$ if and only if the directed cycle is consistent, i.e., the graph is structurally balanced. {\color{black}This implies when $x_1(k-2)$, a previous opinion by agent $v_1$, returns to itself, $v_i$ agrees with what it had previously stated. Otherwise, it contradicts itself, leading to a conflict in perception.}






\textit{Example 2:} Consider the weak cycle in Fig 1 (b). Letting $|a_{ij}|=1$ and $\kappa_i=1/2$ and according to \eqref{DTI1}, one has
\begin{align}
x_1(k+1)&=\frac{1}{2}\big( 1\angle\theta_{12} \big)x_2(k)+\frac{1}{2}\big( 1\angle\theta_{13} \big)x_3(k)\label{x1kk}\\
x_2(k)&=\frac{1}{2}x_2(k-1)\label{x2kk}\\
x_3(k)&=\frac{1}{2}x_3(k-1)+\frac{1}{2}( 1\angle\theta_{32} \big) x_2(k-1)\label{x3kk}
\end{align}
Substituting \eqref{x2kk} and \eqref{x3kk} in to \eqref{x1kk}, one has
\begin{equation*}\label{x1kksum}
\begin{split}
 x_1(k+1)&=\frac{1}{4}\big( 1\angle\theta_{12} \big)x_2(k-1) + \frac{1}{4}\big( 1\angle\theta_{13} \big)x_3(k-1)\\ 
 &+ \frac{1}{4}\big( 1\angle(\theta_{13}+\theta_{32}) \big)x_2(k-1).
\end{split}
\end{equation*}

In the above equation, $1\angle\theta_{12}=1\angle(\theta_{13}+\theta_{32})$ if and only if the weak cycle is consistent, i.e., the graph is structurally balanced. {\color{black}This implies that agent $v_1$ has the same perception of the opinion $x_2(k-1)$, whether it comes directly from agent $v_2$ or is reported by agent $v_3$.}


\begin{figure}[!t]
\centering
\includegraphics[width=2in]{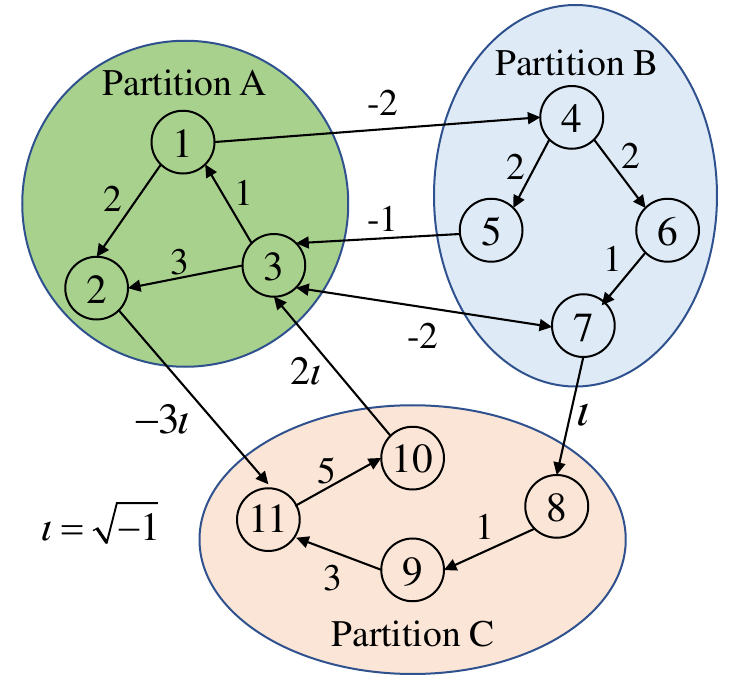}
\caption{Tri-partite graph $\mathcal{G}(A_3)$.}
\label{fig:Tripartite_Graph}
\end{figure}

\section{Simulation}

\begin{figure}[t]
\centering
\subfigure[]{\label{fig:Complex}
\begin{minipage}[t]{0.2\textwidth}
\includegraphics[width=1\textwidth]{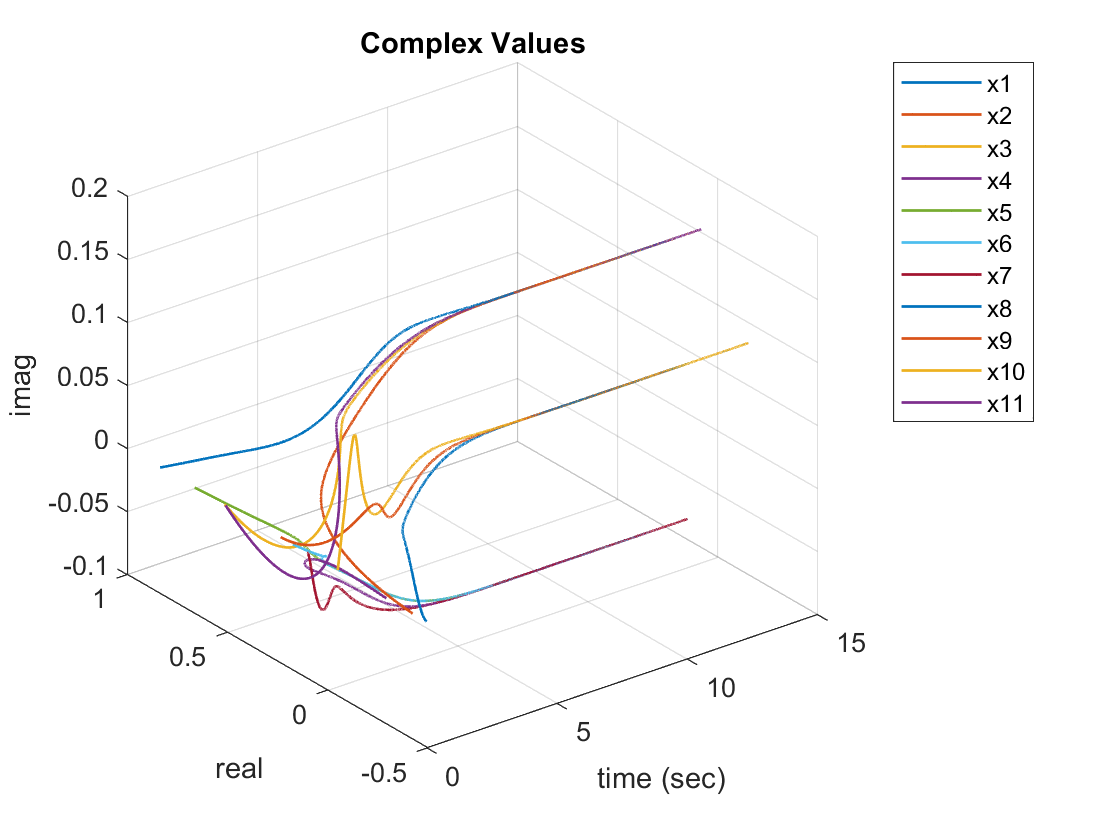}
\end{minipage}
}
\subfigure[]{\label{fig:Real}
\begin{minipage}[t]{0.2\textwidth}
\includegraphics[width=1\textwidth]{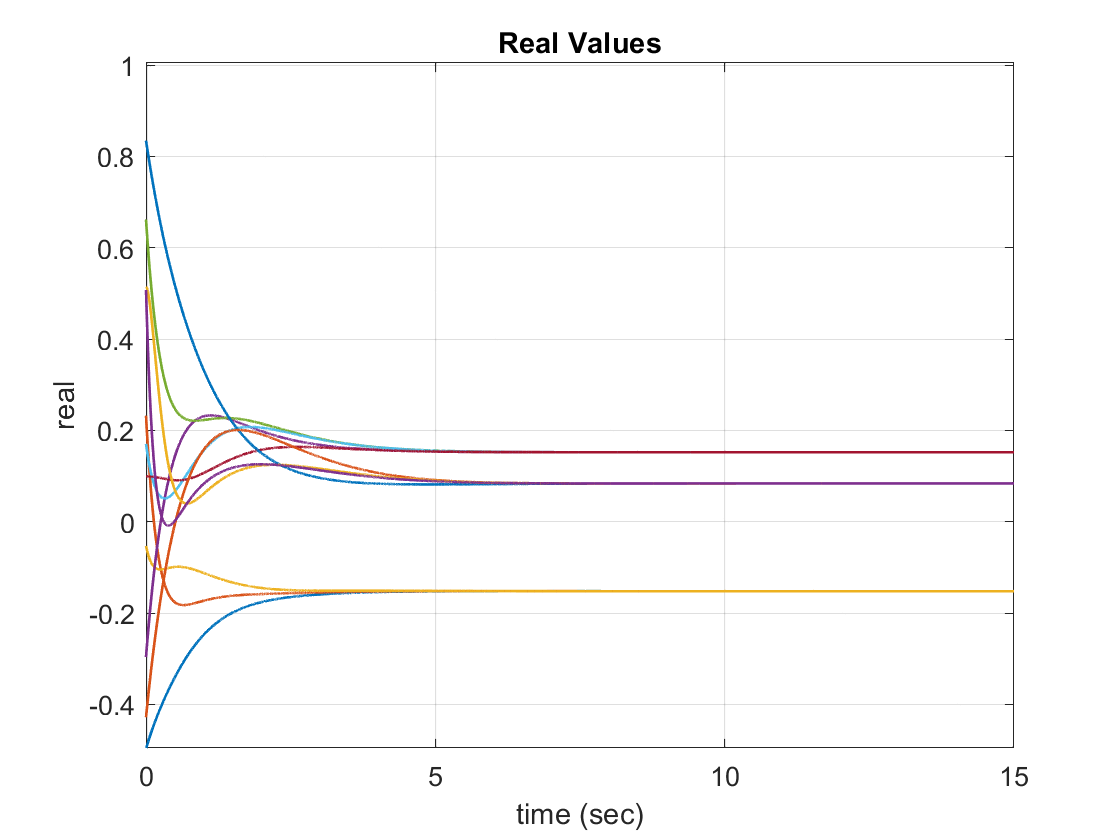}
\end{minipage}
}
\subfigure[]{\label{fig:Imanginary}
\begin{minipage}[t]{0.2\textwidth}
\includegraphics[width=1\textwidth]{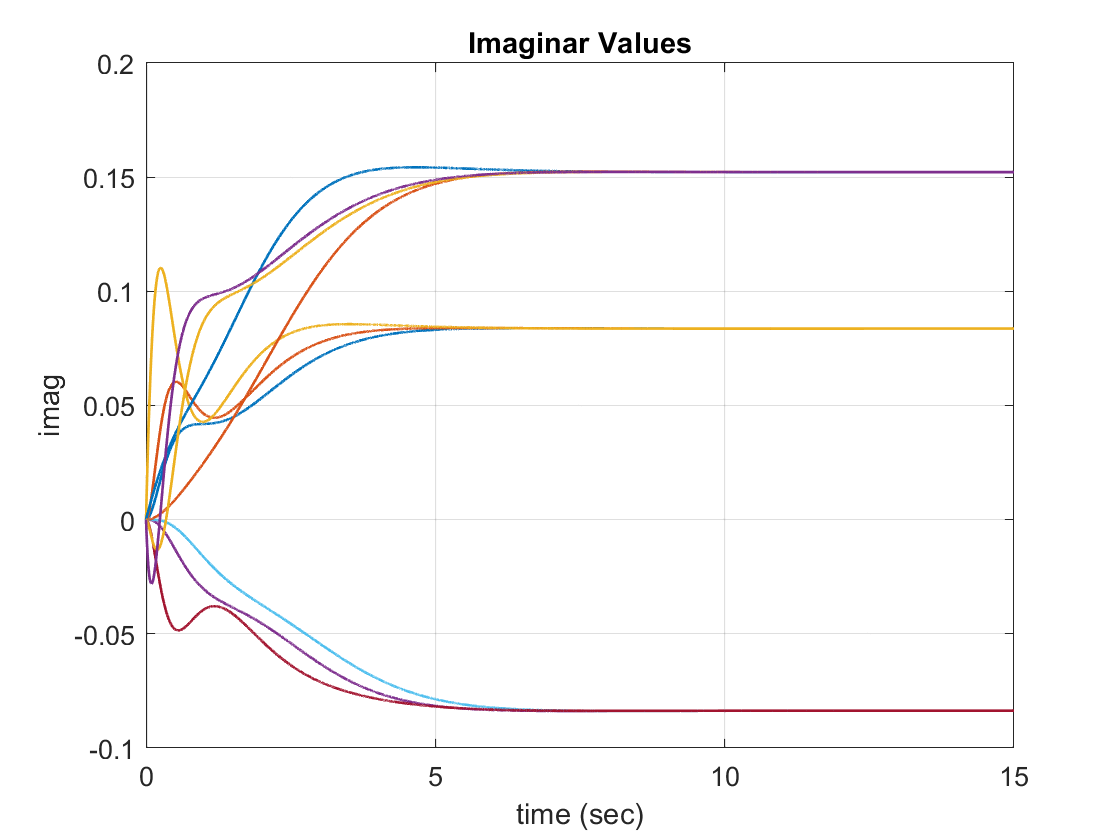}
\end{minipage}
}
\subfigure[]{\label{fig:Absolute}
\begin{minipage}[t]{0.2\textwidth}
\includegraphics[width=1\textwidth]{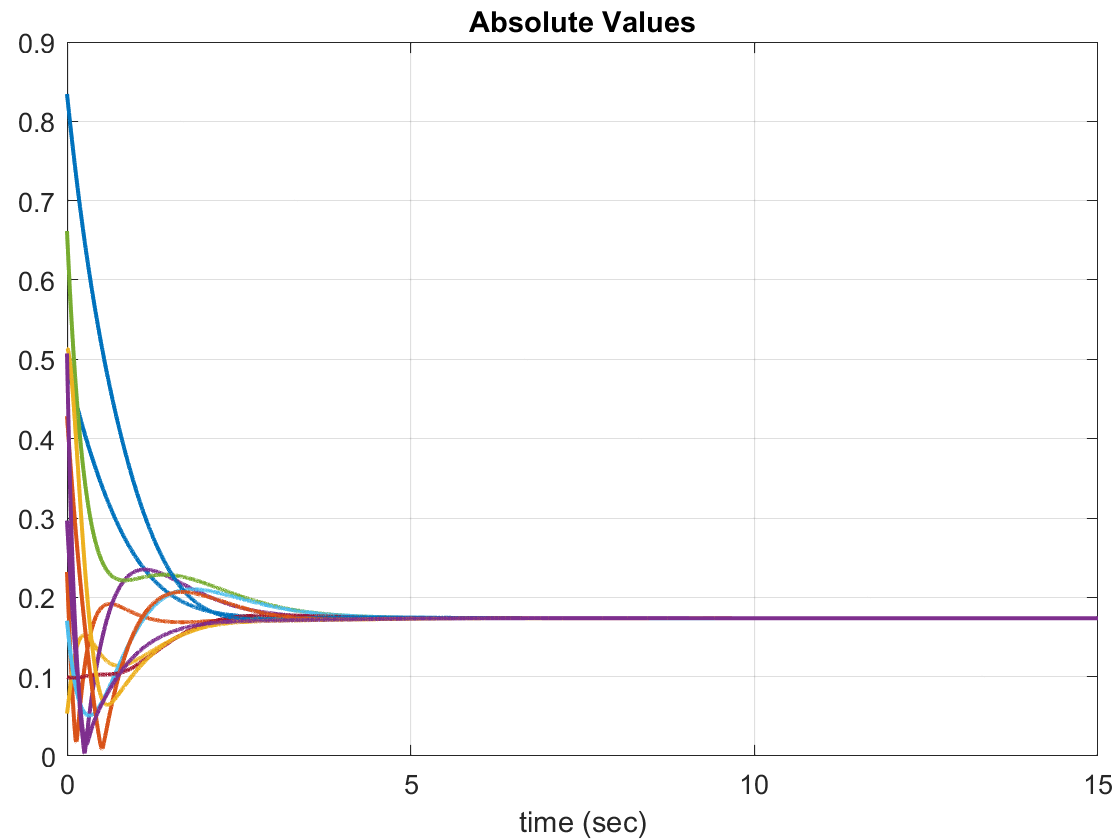}
\end{minipage}
}
\caption{Simulation results in Example 1. (a) Complex values. (b) Real Values. (c) Imaginary Values. (d) Absolute values.}
\end{figure}

\begin{figure}[t]
\centering
\subfigure[]{\label{fig:150Complex}
\begin{minipage}[t]{0.2\textwidth}
\includegraphics[width=1\textwidth]{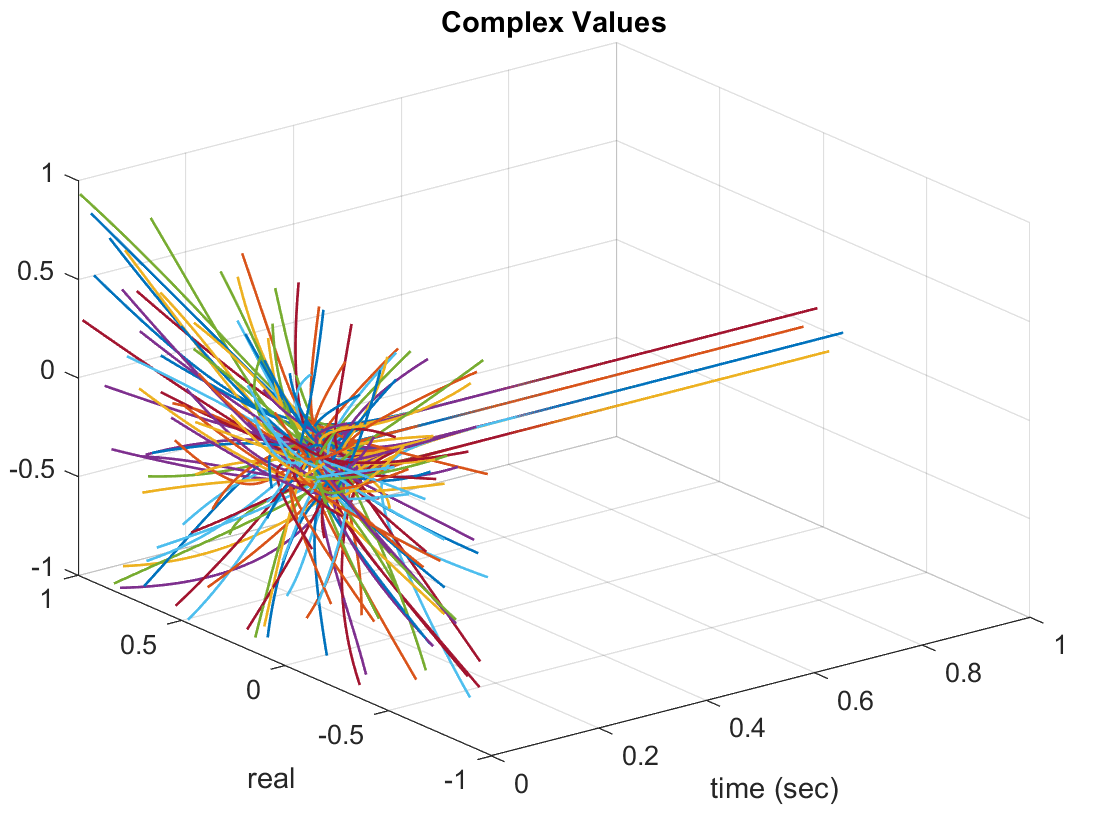}
\end{minipage}
}
\subfigure[]{\label{fig:150Real}
\begin{minipage}[t]{0.2\textwidth}
\includegraphics[width=1\textwidth]{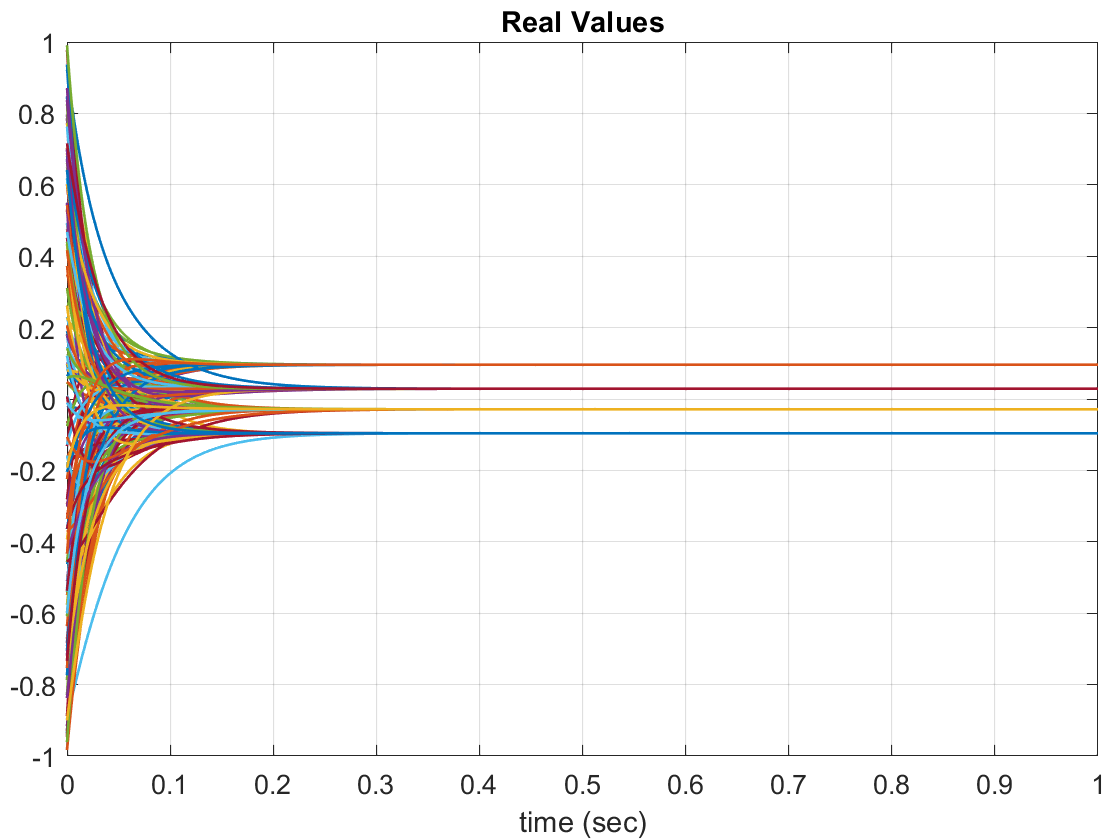}
\end{minipage}
}
\subfigure[]{\label{fig:150Imanginary}
\begin{minipage}[t]{0.2\textwidth}
\includegraphics[width=1\textwidth]{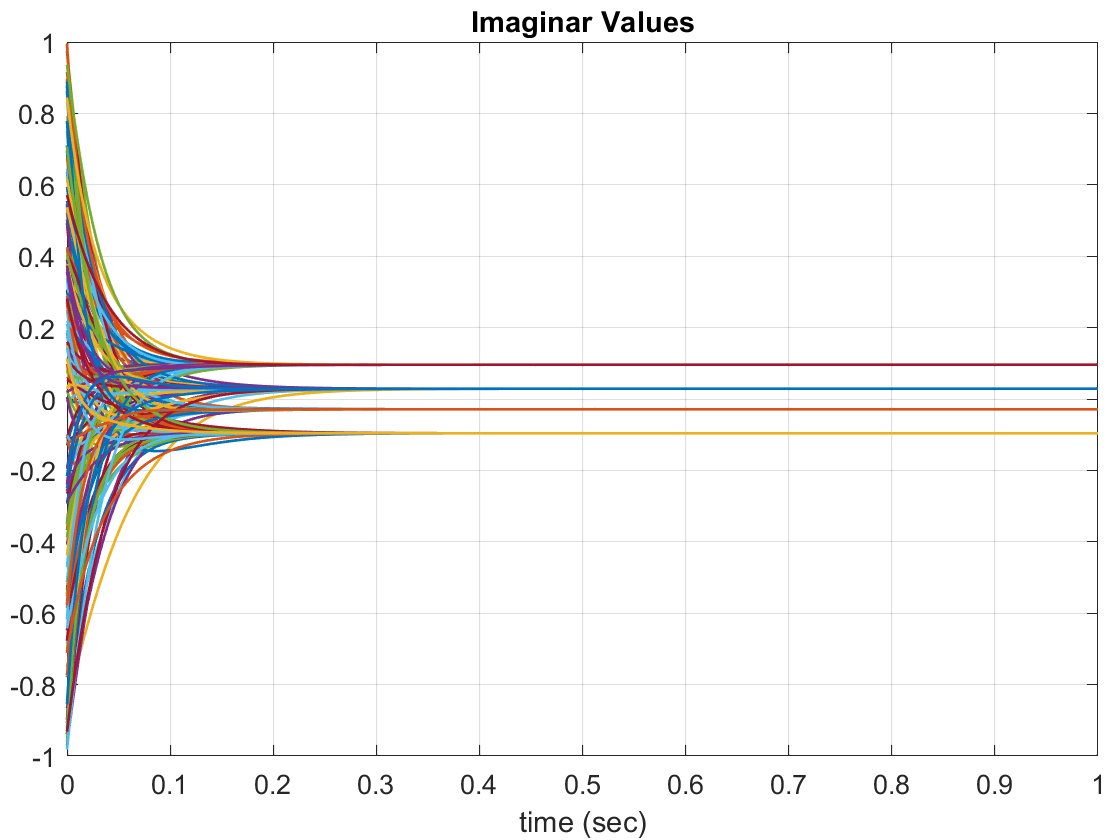}
\end{minipage}
}
\subfigure[]{\label{fig:150Absolute}
\begin{minipage}[t]{0.2\textwidth}
\includegraphics[width=1\textwidth]{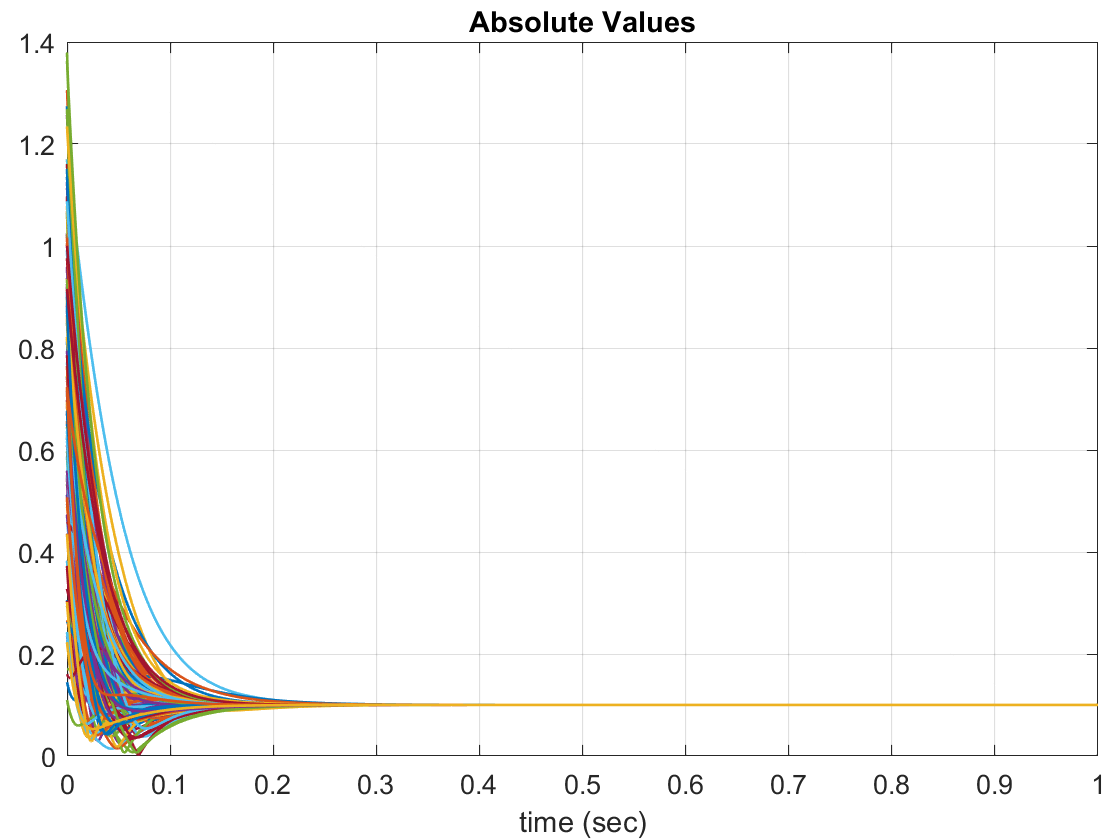}
\end{minipage}
}
\caption{Simulation results in Example 2. (a) Complex values. (b) Real Values. (c) Imaginary Values. (d) Absolute values.}
\end{figure}


\emph{Example 3:} Consider the first-order multi-integers \eqref{CTintegrator} with the local voting protocol \eqref{CTu1} over the graph $\mathcal{G}(A_3)$ in Fig.3. Initialize the states by randomly choosing values from the real interval $[-1,1]$. The simulation results are displayed in Fig.3 (a)-(d), which are the complex, real, imaginary, and absolute values of the states of the first-order multi-integer, respectively. It is shown in Fig. 3 (a)-(c) that the agents in the three partitions of $\mathcal{G}(A_3)$ converge to three separate complex values, respectively. Fig. 3 (d) shows all the agents converge to the same absolute value.

\emph{Example 4:} Consider the multi-integer systems in \eqref{CTintegrator} with protocol \eqref{CTu1} of 150 nodes over a random graph with average connectivity of 0.1 (i.e., the probability is 0.1 that there exists an edge from node $v_j$ to node $v_i$). Each node is randomly assigned with one of the four signatures chosen evenly within $(-\pi,\pi]$ so that four-partite consensus is expected if the graph is structurally balanced. To ensure structural balance, the argument of each edge weight is set to be the difference of the signatures of the two nodes, according to \textit{definition 1}. Set the modulus of the edge weights randomly between 1 to 5. The simulation results are illustrated in Fig. 4 (a)-(d). It is shown that a four-partite consensus is achieved, where the agents converge to four different complex values with the same absolute value.

\section{Conclusion}

In this paper, we studied the multi-partite consensus problem for multi-agent systems, for which we extended the definition of structural balance to complex weighted graphs by introducing signatures to represent the {\color{black}values and beliefs} of the agents. We provided guidelines for constructing a structurally balanced graph. For a given graph without the knowledge of signatures, we developed criteria for checking structural balance, and derived a closed-form solution of the eigenvector $\bm{\zeta}$. This allows for the tools from graph theory, which were initially designed for nonnegative real graphs, to be used with structurally balanced complex weighted graphs.
 Necessary and sufficient conditions were derived for multi-agent systems to achieve multi-partite consensus.
It was shown that ordinary consensus and bipartite consensus are special cases of multi-partite consensus. {\color{black}As a real-world application, complex weighted graphs were utilized to model social networks with multifaceted relationships. It was illustrated that the interactions between agents over a structurally balanced graph promote multi-polarization of opinions.}

\ifCLASSOPTIONcaptionsoff
  \newpage
\fi

\bibliographystyle{IEEEtran}
\bibliography{CWGMPC}

\end{document}